\documentclass[10pt,oneside]{amsart}
\usepackage[utf8]{inputenc}
\usepackage[margin=1.0in]{geometry}
\usepackage{amsmath}
\usepackage[T1]{fontenc}
\usepackage{graphicx}
 
\usepackage{setspace}
\usepackage{listings}
\usepackage{color}
\usepackage[authoryear,round]{natbib}
\renewcommand{\cite}{\citep}
\usepackage[hidelinks]{hyperref}
\usepackage{wrapfig}
\usepackage[autostyle,english=american]{csquotes}

\makeatletter
\def\section{\@startsection{section}{1}%
\z@{.7\linespacing\@plus\linespacing}{.5\linespacing}%
{\normalfont\bfseries}}
\makeatother

\renewcommand{\cite}{\citep}

\title{Rare species advantage in Antarctic Lakes}

\begin{document}

\maketitle

\begin{centering}
Emily Reynebeau, Cristina Takacs-Vesbach, Davorka Gulisija, and Mitchell Newberry\\
{\small Department of Biology, University of New Mexico, Albuquerque, NM 87131}\\
\end{centering}

\begin{abstract}
The maintenance of diversity in complex ecological communities despite
unpredictable dynamics and competitive exclusion is thought to require
continual influx of new species or competitive advantages that accrue as
species become rare. We examine isolated planktonic microbial communities under
permanent ice cover in Antarctic lakes, recording prokaryotic abundance across
9 communities, 11 years, 30~m of depth, and thousands of species in the McMurdo
LTER. We quantify rare species advantage by modeling community dynamics under
frequency-dependent selection. We find persistent diversity and pervasive
negative frequency dependence with limited immigration and turnover. While
ecology and evolutionary sciences have long debated whether diversity is
maintained selectively, we measure selection over a $10^4$-fold range of
abundance in naturally coevolving communities and implicate rare species
advantage.
\end{abstract}

\medskip

Among the longest-standing puzzles in ecology is the paradox of the
plankton---the persistent diversity of species despite the principle of
competitive exclusion \cite{hutchinson1961paradox}. The simplicity and
generality of competitive exclusion has motivated a search for equally simple
and general principles to explain the ubiquity of diversity and coexistence in
nature \cite{chesson2000general,hubbell2001unified}.  Pairwise species
coexistence does not generalize to complex ecosystems \cite{barabas2016effect},
yet neither does pairwise competitive exclusion
\cite{kerr2002local,chang2023emergent}.  Stability in complex ecological
dynamics depends on higher-order species interactions that are potentially
fragile and difficult to predict \cite{may1973stability}, suggesting that
multispecies coexistence hinges on nuanced rules of community assembly
\cite{goldford2018emergent} or coevolutionary history
\cite{legac2012ecological,doebeli2011adaptive}. Neutral theories, by contrast,
explain diversity as the product of continuous influx from mutation, speciation
and immigration, without stabilizing mechanisms
\cite{macarthur1967theory,kimura1985neutral,hubbell2001unified}.

Ecology and evolutionary sciences have converged on rare species advantage as
an essential feature of stabilizing mechanisms. Balancing selection, mutual
invasibility, negative density dependence and modern coexistence theory all
involve advantages to rare types
\cite{grainger2019invasion,johnson2012conspecific,chesson2000general}.
Specialization of natural enemies on common species \cite{janzen1970herbivores}
or variability in recruitment favoring rare species
\cite{warner1985coexistence} are thought to maintain the diversity of tropical
trees and reef fishes \cite{tanner2009community,kalyuzhny2023pervasive}. Yet
these proposals have been difficult to test quantitatively, engendering debates
on niche versus neutrality across fields.  Both neutral and density-dependent
theories are consistent with diversity inventories
\cite{volkov2005density,chisholm2010niche}, indicating that diversity
\textit{per se} offers weak evidence of its origins. Detecting negative density
dependence in forests has been fraught with confounds
\cite{johnson2012conspecific,broekman2019signs,detto2019bias,hulsmann2024latitudinal}
and seems to require long-term dynamic data
\cite{kalyuzhny2023pervasive,hulsmann2024latitudinal}. Rare species advantage
is therefore widely assumed, but measurements in diverse communities are
scarce.

We measure rare species advantage in terms of frequency-dependent selection
(FDS) in microbial time series. FDS quantifies how competitive ability depends
on abundance in commensurate terms across ecology and evolution, where
selection is the differential growth rate of a focal species relative to the
community average
\cite{newberry2022measuring,christie2023negative,chang2023emergent}.  If
frequency dependence is negative, competitiveness declines with increasing
abundance, indicating rare species advantage. New methods exist to measure FDS
from abundance time series of whole communities \cite{newberry2022measuring},
while the revolution of DNA barcoding using the 16S ribosomal RNA gene has
enabled time courses of microbial abundance along development, disease
progression and environmental change \cite{gilbert2012defining}.

The short lifetimes of microbes allow tens of thousands of generations to be
observed on laboratory timescales \cite{good2017dynamics}. Laboratory
experiments have indeed observed coexistence in naturally-derived communities
in artifical environments \cite{chang2023emergent} and the spontaneous
evolution of coexisting strains of bacteria
\cite{rozen2000long,legac2012ecological} and their viral symbionts
\cite{meyer2016ecological}. Natural microbial time series, however, are
typically dominated by species flux and environmental or host-associated
exogenous forcing that swamp intrinsic competitive dynamics
\cite{fuhrman2006annually,campbell2011activity,gilbert2012defining}.

\begin{figure}[t]
\makebox[\textwidth][c]{\includegraphics[width=180mm]{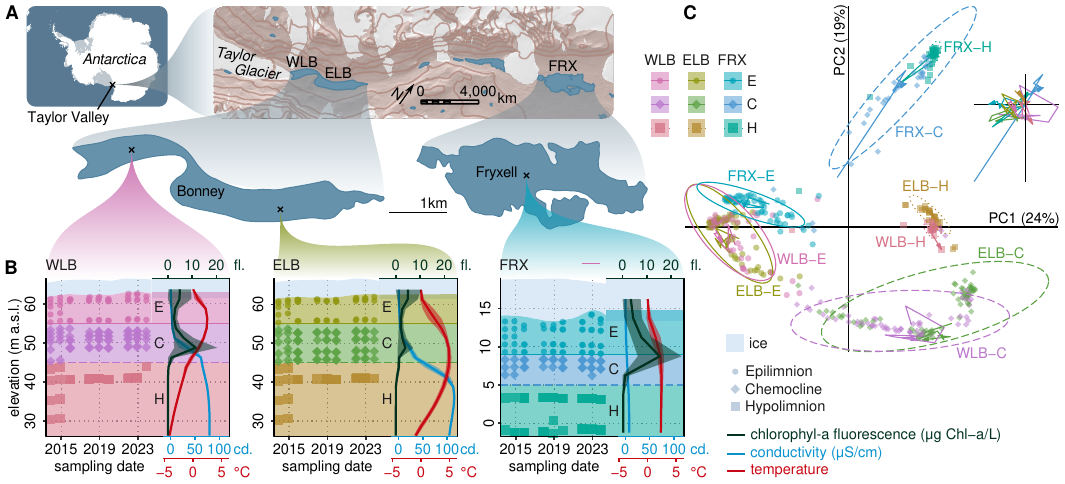}}
\caption{\textbf{McMurdo Dry Valley lakes harbor distinct microbial communities
stratified by depth.} (A) MDV lakes Bonney and Fryxell are located in the
Taylor Valley of Southern Victoria Land in East Antarctica. (B) 16S ribosomal
gene sampling spanned 15-30 m of depth in three lakes over 11 summers. Depth
profiles of temperature, salinity and chlorophyll fluorescence correspond to
stable stratification into epilimnion, chemocline and hypolimnion with two
photosynthetic zones. (C) Species composition clusters by lake and depth
stratum on principle coordinate axes (Bray-Curtis PCoA). Time courses of
samples pooled by depth stratum (solid lines) remained within clusters, showing
greater homogeneity across time than across communities (inset: recentered time
series with 2013 value at the origin).}
\label{fig:lakes}
\end{figure}

Glacier-fed lakes of the McMurdo Dry Valleys (MDVs) of Antarctica harbor unique
planktonic microbial communities with long coevolutionary history in a stable
and isolated environment \cite{li2019influence}. We sampled and sequenced
prokaryotic 16S ribosomal genotypes (ribotypes) in three MDV lakes
(Fig~\ref{fig:lakes}A): Lake Bonney's west (WLB) and east (ELB) lobes and Lake
Fryxell (FRX). These hyperstable, meromictic lakes of liquid water are
permanently covered in 3 to 6 meters of ice that shelter communities against
migration, allochthonous input and seasonal, wind- or temperature-induced
mixing. Mixing times exceed 10,000 years \cite{spigel1996evolution} and sinking
rates are negligible \cite{takacs1998bacterioplankton} supporting limited
dispersal in the microplankton. Bacteria are by far the most numerous cells by
up to three orders of magnitude \cite{takacs1998bacterioplankton} and comprise
up to 50\% of lake biomass. The remainder includes uni- and multicellular
eukaryotic microplankton but no higher plants or animals \cite{goldman1967two}.
Minimal biotic \cite{li2019influence} and abiotic \cite{spigel2018physical}
flux means that changes in species composition can be attributed to endogenous
dynamics near steady state rather than exogenous forcing.

This region of Antarctica has escaped warming, with cooling until 2001 and
stable air temperature and solar radiation in recent decades
\cite{obryk2019prediction}. Liquid moats form in late summer at lake edges
200-600 m from sampling locations, but harbor distinct communities and mix
minimally with the interior \cite{castendyk2015pressure,stone2024mcmurdo}.
However, MDV lakes are now experiencing increased heat flux and rising lake
level, and are predicted to be seasonally ice free by 2050
\cite{obryk2019prediction}, threatening this unique genetic resource.

We collected during austral summers (Nov 4-Jan 8) of 2013-2015, 2017-2019 and
2021-2023, producing three intervals of three consecutive years and over 52
million reads sequenced on Illumina NextSeq 2000. We used Niskin bottles to
collect water along 19.7-35 m depth transects at 8-25 points per lake per year,
totaling 390 sample points (Fig~\ref{fig:lakes}B). We identified 10,776
ribotype species after DNA sequence data quality control
\cite{callahan2016dada2}, distinguished as amplicon sequence variants (ASVs).
We conservatively report total identified species and otherwise verified
consistency of results across alternative denoising assumptions
\cite{bardenhorst2022richness}.

\begin{figure}[t]
\makebox[\textwidth][c]{\includegraphics[width=118mm]{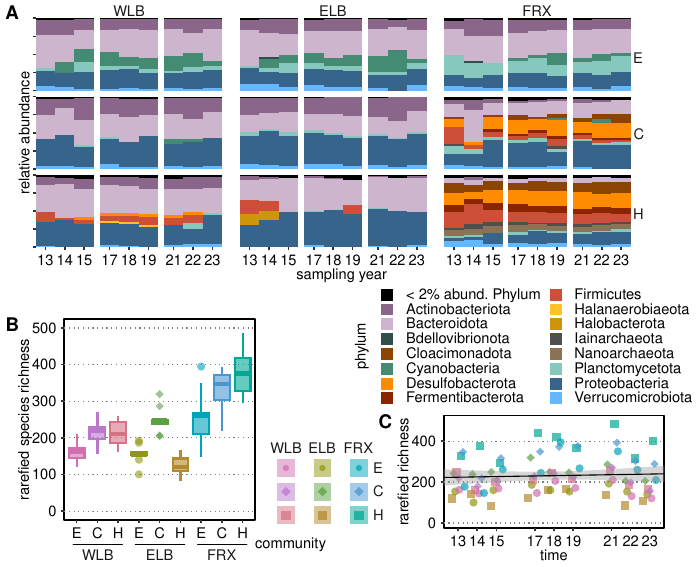}}
\caption{\textbf{Diversity and taxonomic compostition of communities persists
over time}. (A) Phylum-level community composition persists across time. (B)
Rarefied species richness varies by community over a 3.1-fold range (Tukey box
plots of species richness measured at different times). (C) Richness is not
correlated with time.}
\label{fig:div}
\end{figure}

We observed distinct communities and stratification along physicochemical
gradients (Fig~\ref{fig:lakes}C, Fig~\ref{fig:div}A).  The shallow epilimnion
(E), hyperoxic from oxygen exclusion in ice formation and photosynthesis,
contained similar taxa across lakes---primarily Bacteroidota (26-49\% of
individuals) followed by Proteobacteria, Actinobacteriota, Cyanobacteria and
Planctomycetota (Fig~\ref{fig:div}A). A chemocline (C) occurring 10-20 m below
ice supported a second peak of photosynthetic potential (Fig \ref{fig:lakes}B),
where light is scarce but nutrients are abundant. Here, community composition
diverged by lake (Fig~\ref{fig:lakes}C), species richness increased
(Fig~\ref{fig:div}B) and Proteobacteria dominated (14-46\%).  The chemocline of
FRX is more shallow, less acute, influenced by a blend of epi- and hypolimnion
chemistries, and uniquely supported abundant anaerobic (Cloacimonadota,
Fermentibacterota), sulfur cycling (Desulfobacterota), and copiotrophic
(Firmicutes) phyla.  The deep and longest-isolated hypolimnion (H) contained
anoxic, brackish to hypersaline and often sub-freezing water and supported the
most divergent communities between lakes. Firmicutes and the halophilic Archaea
Halobacterota occurred in Lake Bonney, while the brackish H of FRX supported an
active sulfur cycle dominated by Desulfobacterota (11-23\%). Archaea occurred
in all communities, but were typically rare. Of the seven archaeal phyla
identified, only three amount to more than 0.1\% of the community, and three
others only occur in FRX. The most consistently sampled communities of abundant
Archaea were Nanoarchaeota (7-10\%) and Iainarchaeota (2-8\%) in the FRX H.

We pooled samples within each depth range and field season to create time
series of interannual dynamics for nine communities---all possible combinations
of WLB, ELB, FRX and E, C, H. Pooling trades spatial and temporal precision for
frequency resolution and data harmonization---improving spatial and temporal
comparability between time points, lowering noise and providing better
resolution of the frequency of rare species. These time series therefore
capture long-term community dynamics, marginalizing over seasonal variation.

These communities retained diversity and composition across time.  We found 17
species to be ubiquitous in all samples, yet these ubiquitous species
represented a highly variable fraction of each community, ranging from 0.9 to
65\%. Rarefied species richness to 99\% coverage \cite{chao2020quantifying}
differed across communities over a 3.1-fold range (Fig.~\ref{fig:div}B), with
significant pairwise differences between most communities (22 out of 36 Tukey
adjusted $p<0.01$). Richness reflected compositional differences between
communities, with higher richness in FRX and more variation in deeper
communities. Time courses remained within community clusters on PCoA, with no
particular directional change (Fig.~\ref{fig:lakes}C). Rarefied richness was
not correlated with time (Fig.~\ref{fig:div}C, Pearson $\rho$=$0.06$,
$p$=$0.61$) or sample size ($\rho$=$0.10$, $p$=$0.38$), consistent with
diversity near steady state.

\begin{figure}[t]
\includegraphics[width=118mm]{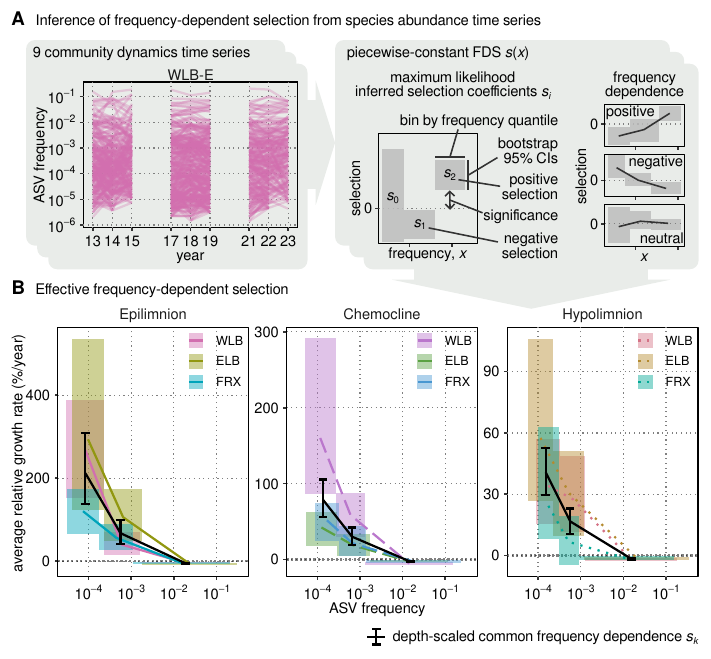}
\caption{\textbf{All communities exhibit negative frequency dependence.} (A) We
infer the effective frequency-dependent selection in each community by fitting
time series of species abundance to a simple model of frequency-dependent
selection by maximum likelihood. (B) Frequency dependence is negative and
retains its shape across lake and depth. A common form of frequency dependence
$s_k$ shared by all communities scaled by a depth-dependent rate constant
$\delta_i$ (black lines) explains 79.8\% of the variance in the inferred
selection coefficients.} 
\label{fig:fdsel}
\end{figure}

We measure FDS following a general method \cite{newberry2022measuring} to fit a
model of frequency-dependent competition to time series using maximum
likelihood (Fig.~\ref{fig:fdsel}A). We parameterize FDS as a piecewise-constant
function $s(x)$ that assigns a selection coefficient $s$ to species at
frequency $x$. The maximum-likelihood selection coefficients for each frequency
range are equal to the average selection within the range, called the effective
FDS, regardless of whether the true dynamics are strictly frequency dependent
\cite{newberry2022measuring}. The effective FDS thus measures community-wide
average association between relative growth and frequency.  The ``effective''
FDS falls short of perfectly matching the invasion criterion, in that a
positive average growth when rare does not imply that each rare species’ growth
rate is positive, but it nonetheless reports community-wide tendency to favor
rare species.  Biases such as regression dilution have affected measurements
through error prone proxies \cite{detto2019bias}. In our case, sampling
variability at low frequency sets the minimum frequency of valid inference,
which we determine through controls for sampling bias
\cite{newberry2022measuring}. 

We find strong, nonlinear, negative FDS in each community
(Fig.~\ref{fig:fdsel}), indicating measurable advantage to species when they
are rare, down to frequency 0.003\%. These 9 inferences incorporate 1,158
species and between 105 and 345 species at any given time. In all cases,
selection is positive in the rarest frequency range and negative in the most
common and excludes zero according to 95\% bootstrap confidence intervals
\cite{newberry2022measuring}.  Common species at frequencies between 0.1 and
35\% decline in frequency on average at rates from $-6.9$ to $-1.2$\%/year,
whereas the rarer species from 0.003-0.2\% increase on average at rates ranging
from $6.7$ to $300$\%/year depending on frequency, lake and depth.

We furthermore find a common shape of $s(x)$ across communities, scaled by an
overall rate that depends on depth. We extract a shared form of $s(x)$ across
all communities through singular value decomposition (SVD) by considering the
inferred selection coefficients $\hat{s}_{ijk}$ for all lakes $i$, depths $j$,
and frequency ranges $k$ as products of best-fit lake-, depth- and
frequency-dependent parameters $\lambda_i$, $\delta_j$ and $s_k$. The shared
frequency-dependent selection $s_k$ by itself explains 49\% of the variance in
$\hat{s}_{ijk}$. Scaling $s_k$ by the depth-dependent rate constant $\delta_j$,
as $\hat{s}_{ijk} = s_k\delta_j + \epsilon$, explains 80\% of the variance
whereas the full product, $\hat{s}_{ijk} = s_k\delta_j\lambda_i + \epsilon$,
explains at most 87\%. No other combination of lake, depth or frequency
dependence explains any comparable fraction. Depth therefore scales the overall
rate of the selective dynamics, with faster dynamics at shallower depth over a
3.3-fold range. We plot the shared form of $s(x)$ as the product $s_k\delta_j$
as bold black lines in Fig.~\ref{fig:fdsel}B. 

\begin{figure}[t]
\includegraphics[width=54mm]{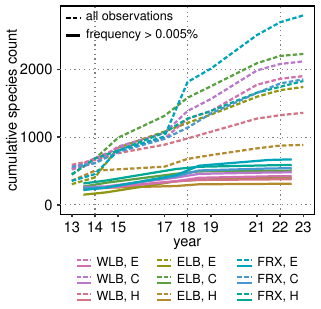}
\caption{\textbf{Saturation in species-time curves indicates limited
immigration and turnover.} During the last third of time series, novel
ribotypes are observed at rates at most 8.5\%/year and at most 1.4\%/year
attain frequencies within the observation window. The long-term rate of
immigration to observable frequency is therefore bounded to at most 1 to 9
novel ribotypes/year across communities.}
\label{fig:turnover}
\end{figure}

Immigration rates are categorically difficult to measure, but we provide a
robust upper bound rate of 1.4\% novel species per known species per year
reaching measurable frequency. Noise sources in presence/absence sampling
include sampling error, variation in sample size, amplicon sequencing error,
uncertainty in read calling and ribosomal mutation \cite{callahan2016dada2}.
Species-time curves, like species-area curves, account for more observations
with greater sampling effort \cite{white2006comparison} and allow some error
correction. Saturation in cumulative species count distinguishes high,
short-term rates of novel observations due to undersampling from lower
long-term rates due to immigration, mutation and amplicon sequencing errors. We
observe partial saturation in all time series (Fig.~\ref{fig:turnover}). Novel
species accrue during the first third of time series at higher rates ranging
from 21 to 440 ribotypes/year (4.2-109\%/year), versus lower rates from 9 to
189 ribotypes/year or 1.0 to 8.5\%/year during the final third (ANOVA period
$p$=0.002). FDS inference incorporates only species that reach a reliably
measurable frequency at multiple consecutive time points, which further
excludes amplicon sequencing errors and reduces sampling noise. Cumulative
species incorporated into the inference accrue at 4 to 57 ribotypes/year in the
first third and 1 to 9 ribotypes/year or 0.2-1.4\%/year in the final third
(ANOVA period $p$=0.006). This maximum long-term rate of 1.4\%/year still
includes contributions from ribosomal mutation and undersampling, and so
provides an upper bound on the rate at which new species immigrate and achieve
measurable frequency.

Top list turnover rates are theoretically independent of sample and population
size and informative of the underlying dynamics \cite{acerbi2014biases},
provided that the lists represent a small fraction of the total number of
detected species \cite{eriksson2010bentley}. We measure turnover in lists of
top 25 most abundant species, representing 11\% of the species observed in
typical samples. Mean number of ribotypes entering or leaving top-25 lists per
year varied from 2.4 to 6.9 across time series and associated significantly
with depth (ANOVA depth effect $p$<0.001, lake effect $p$=0.81). Turnover in E
and C communities (5.5-6.9 ribotypes per year) occurred at 1.9 times the rate
of H (2.4-3.6 ribotypes per year). The effect does not depend on the length of
the top list, such as top 35 or top 50, though the particular rates do, as
expected \cite{eriksson2010bentley}.

Rates of overall selection as well as top list turnover decrease with depth, by
factors of 3.3 and 1.9 respectively, suggesting faster overall community
dynamics at shallower depth. This finding is consistent with higher bacterial
production and biomass turnover coinciding with the two chlorophyll peaks in
the epilimnion and chemocline
\cite{takacs1998bacterioplankton,burnett2006environmental}. More complex
interspecific interaction have also been observed in shallow phototrophic
communities \cite{li2019influence}, potentially due to predation, mutualism,
and other interactions \cite{patriarche2021year}. 

Diversity at steady state does not depend on the overall rate, but rather on
the shape of frequency-dependent selection and relative rates of selection,
speciation, immigration and stochastic drift. We find high diversity at great
depth, where there is slower selection and turnover, as well as in Lake
Fryxell, where there is higher overall productivity, moat size and stream input
\cite{takacs1998bacterioplankton}. Negative frequency dependence therefore does
not rule out influence from geology, migration and natural history.  Similarity
between communities (Fig.~\ref{fig:lakes}C) generally reflects the degree of
physical connectivity between lakes and depths, geochemical composition and
geologic history (Fig.~\ref{fig:lakes}A) \cite{lyons2000importance}, while
photosynthetic community composition might reflect environmental similarity or
metacommunity input from glacial streams \cite{takacs2001bacterial,
lyons2003surface}, wind \cite{sabacka2012aeolian, schulte2022blowin} or the
soil-moat-lake interface \cite{stone2024mcmurdo}. Both the highest diversity
and the highest distinctiveness between lakes occur in the deepest communities,
which have remained isolated by halostratification since a dry down and
evapoconcentration event occurring approximately 15 ka BP
\cite{hall2017constraining}.

The pervasive pattern of negative frequency-dependent selection across
communities suggests general features of competitive dynamics that favor rare
species. Negative frequency dependence clearly contradicts neutrality and
competitive exclusion in favor of selective mechanisms that maintain diversity.
Physical isolation and observed limited immigration provide further evidence
that these selective mechanisms are necessary for diversity to persist for
hundreds or thousands of years. These findings in isolated coevolving microbial
communities therefore inform long-term debates on the roles of selective
forces, coexistence mechanisms, and coevolutionary history in maintaining
diversity.

Negative effective frequency dependence is an emergent property of complex
dynamics. These findings are therefore consistent with a broad category of
hypothesized community dynamics, such as exclusion of the fittest
\cite{shibasaki2021exclusion}, kill the winner \cite{thingstad2000elements},
boom-bust cyles \cite{doebeli2021boom,newberry2022measuring}, diversity waves
\cite{maslov2015diversity}, and storage effects \cite{chesson2000general}---all
of which provide some benefit to rare species. MDV lake bacterial populations
are known to support high viral loads
\cite{robinson2024antarctic} thought to create negative
density-dependent mortality \cite{thingstad2000elements} while oceanic bacteria
show long-term cycles in abundance \cite{campbell2011activity}. Our findings
support that these dynamics promote diversity in real systems.

Investigating FDS in further systems may reveal general tendencies of species
interaction and evolution to foster and protect diversity. We caution, however,
that negative FDS in MDV lake communities may be a product of their
coevolutionary history, which may disappear as predicted warming and transition
to seasonal ice allow increased influx \cite{li2019influence}. Coexistence has
been both theorized \cite{doebeli2011adaptive} and observed
\cite{legac2012ecological} to result from coevolution. Whether assemblages with
weaker shared evolutionary history, such as invasive species or the products of
anthropogenic dispersal, also create stabilizing interactions remains a key
question for conservation. The biodiversity crisis is widely appreciated in
terms of species loss, yet loss of species interactions that stabilize
coexistence might be just as important.

\par\bigskip\par
\paragraph*{Acknowledgments}
The University of New Mexico Comprehensive Cancer Center provided access to the
Analytical and Translational Genomics Shared Resources for Illumina
next-generation sequencing. We appreciate eleven years of fieldwork from the
MCM-LTER field teams. We would also like to thank the Antarctic Support
Contractors and Air Center Helicopters for providing operational support in the
field, without which the quality of Antarctic research performed today would
not be possible.
\par\medskip\par
\paragraph*{Funding}
This work was funded by the National Science Foundation (NSF) OPP-1115245,
1637708, and 2224760 McMurdo Dry Valleys Long Term Ecological Research Project
as well as NSF OPP-1937627 awarded to CTV which provided support for ERR. ERR
also received support from the University of New Mexico Biology Department. A
portion of this publication was developed for the dissertation of ERR. MN was
funded by DG's startup at University of New Mexico.
\par\medskip\par
\paragraph*{Author contributions} Concept: MN,CT; Sampling: CT,ER; Sequencing:
ER; Theory: MN,DG; Data analysis: MN,ER; Supervision: MN,CT; Figures: MN,ER;
Writing: MN,ER,CT,DG
\par\medskip\par
\paragraph*{Competing interests} There are no competing interests to declare.
\par\medskip\par
\paragraph*{Data and materials availability}
Original sequences and metadata will be available on NCBI and archived in the
Environmental Data Initiative (EDI),
(DOI: 10.6073/pasta/cdb05d6d9e0138135b8e7b324ad93181).\\
Water column physical and chemical data are available on
\url{https://mcm.lternet.edu/} and archived in the EDI including:\\
ice thickness (DOI: 10.6073/pasta/e72dc49d774796767884c535e864c915),\\
lake level (DOI: 10.6073/pasta/649e1e54f663e8077f6ca96352e703ba),\\
conductivity and teperature (DOI: 10.6073/pasta/92d06937855a37329bec519c7dd3cbd5),\\
and fluorecence (DOI: 10.6073/pasta/19d2d4d306e97782a43951be78822aa2).\\
The full data analysis pipeline in R and OCaml will be available on\\
\url{https://github.com/mnewberry/mcmlter-16s-fds}

\end{document}